\documentclass[%
 reprint,
 amsmath,amssymb,
 aps,
]{revtex4-2}
\usepackage{braket}
\usepackage{diagbox}
\usepackage{xcolor}
\usepackage{graphicx}
\usepackage{soul}
\usepackage{dcolumn}
\usepackage{bm}

\begin{document}

\preprint{APS/123-QED}

\title{A context-aware gate set tomography characterization\\ of superconducting qubits}

\author{Ahmed Abid Moueddene}
\email{A.A.Moueddene@tudelft.nl}
 \affiliation{QuTech, Delft University of Technology
Delft, The Netherlands}
\affiliation{Quantum Computer Engineering Dept, Delft University of Technology
Delft, The Netherlands.}
\author{Nader Khammassi}%
 \email{ Nader.Khammassi@intel.coml}
\affiliation{Intel Labs, Intel Corporation, Hillsboro, Oregon, USA
}%
\author{Sebastian Feld}%
 \email{ S.Feld@tudelft.nl}
  \affiliation{QuTech, Delft University of Technology
Delft, The Netherlands}
\affiliation{Quantum Computer Engineering dept, Delft University of Technology, Delft, The Netherlands.}

\author{Said Hamdioui}
 \email{S.Hamdioui@tudelft.nl}

\affiliation{Quantum Computer Engineering dept, Delft University of Technology, Delft, The Netherlands.}%

\date{\today}

\begin{abstract}
The efficiency of Quantum Characterisation, Verification, and Validation (QCVV) protocols highly hinges on the agreement between the assumed noise model and the underlying error mechanisms. As a matter of fact, errors in Quantum Processing Units (QPUs) incorporate various aspects of context-dependability which are overlooked by the majority of the commonly used QCVV protocols. As QCVV protocols are indispensable when it comes to characterizing and evaluating quantum operations, there is a serious need for a detailed characterization taking into account such aspects. In this work, we address these shortcomings by designing a context-aware version of the gate set tomography (GST) protocol. Our experiment selection approach is based on a polynomial quantification of the accumulation of errors within the designed circuits. Using simulated QPUs, we show that this technique enables a characterization with an inaccuracy reaching $10^{-5}$. Furthermore, we use our proposed protocol to experimentally infer context-dependent errors, namely crosstalk and memory effects, in a publicly accessible cloud-based superconducting qubits platform. Our results show that when the GST is upgraded to include such features of context-awareness, a large coherence in the errors is observed. These findings open up possibilities of drastically reducing the errors within the currently demonstrated QPUs.

\end{abstract}

\maketitle

\section{Introduction}

 A quantum state is by essence destructed when measured \cite{qiandqc}. Therefore, real-time monitoring of the computational states of QPUs while preserving their quantum aspect is generally unlikely. In combination with being subject to numerous noise mechanisms, this fact implies that the creation of quantum states and the different operations applied to them can only be evaluated indirectly. In quantum computing, these practices are commonly known as QCVV protocols. They consist of sampling  sequences of operations and converting the observed measurements into meaningful parameters. Designing efficient QCVV protocols is a major challenge in this field, as they are vital for understanding noise mechanisms \cite{Dehollain_2016,erhard}, calibrating control signals  \cite{HW1,HW2}, developing noise-aware compilers \cite{Murali2020}, and providing realistic noise models for simulation backends  \cite{UMC}. 
 
There exist a wide range of QCVV protocols designed for targeting a variety of noise mechanisms  \cite{Sarovar2020,Helsen2019,pygstynew,PhysRevLett.77.4281,QSE,QST,qpt,doi:10.1080/09500349708231894,Xin2019,rb,rbleak,ctrb,Palmieri2020,Govia2020,1310.4492,gstintro,GST2q,2009.07301,erhard}. These protocols exhibit an interplay between the number of extracted parameters, the expected accuracy, the number of the measured sequences, and the awareness of systematic errors. For instance, Randomized Benchmarking (RB) protocol is used to infer the depolarizing rate of quantum gates  \cite{rb}. This protocol owes its high accuracy to its output corresponding to a single parameter (the error rate) and its awareness to State Preparation And Measurement (SPAM) errors. other examples are  quantum state tomography  \cite{Pauli1980,PhysRevLett.77.4281,QSE,QST} and  quantum process tomography \cite{qpt,doi:10.1080/09500349708231894}. They are used to respectively infer quantum states and quantum operations. These protocols provide a more detailed characterization as their outputs are density matrices and quantum channels. However, unless properly addressed  \cite{Palmieri2020,Govia2020}, their standard forms have moderate accuracy, as they are oblivious to systematic errors and based on sampling a relatively small number of sequences. As an alternative, the Gate Set Tomography (GST) protocol provides a more accurate and detailed characterization since it takes into account and evaluates SPAM errors and relies on sampling a larger number of sequences \cite{1310.4492,gstintro,GST2q,2009.07301}. Although the requirement of measuring a large number of sequences comes with a high time complexity (fridge time), it was demonstrated that involving the GST protocol in the calibration routines leads to a considerable reduction in the underlying operational errors  \cite{Dehollain_2016,PhysRevApplied.15.014023}. In other works, data delivered by measuring the GST sequences were used to infer deliberately engineered temporally correlated noise processes  \cite{Mavadia2018} and context-dependent errors  \cite{PhysRevX.9.021045,Proctor2020}. However, the state of the art suffers from the lack of GST protocols adapted to capturing sophisticated forms of errors, namely crosstalk in large arrays of qubits(beyond three qubits) and memory effects. These noise mechanisms were not captured by the original GST protocol
 
 To characterize crosstalk errors in a set of qubits; how a given operation on a target qubit is affected given the operations simultaneously applied to the neighboring qubits, one can conceive two approaches involving the context agnostic GST. The first one is to perform a multiqubit GST on the whole set then trace out the neighboring qubits. This approach is impractical as the operators describing multiqubit gates are massive for sets containing more than 2 qubits. The second alternative is to perform multiple rounds of single-qubit GST on the target qubit. In each round, the target gates are accompanied by a different configuration of gates on the neighboring qubit which enables inferring crosstalk. In this paper, we introduce a context-aware version of the  GST protocol. It systematically takes into account spatial correlations which allows inferring the target operations in a multiplicity of scenarios within a single round. Therefore, it reduces considerably the time complexity that comes with the requirement of running multiple rounds of the latter approach while it tackles the impracticality of the first one. Additionally, it also takes into account spatial correlations which allows characterizing memory effects. Therefore, the proposed protocol is a step towards enabling a rigorous assessment of the erroneous behavior of QPUs. In short, the main contributions of this paper are:
\begin{itemize}

    \item The development and the validation of a  new version of the GST protocol based on the refinement of the sequence selection algorithm, allowing a high error characterization accuracy.
    
      \item The differentiation of the proposed version by introducing the context-awareness, enabling the characterization of crosstalk errors and memory effects.

    \item The demonstration of the protocol on   Quantum-Inspire's Starmon-5 chip \cite{inspire}; the results reveal a relevant context-dependency in the error mechanisms affecting this chip with considerably large reversible parts.
    
    \end{itemize}

Suppose one possesses a QPU offering the possibility to run a target gate set composed of state preparation, a set of $n$ gates $\{G_1,G_2...,G_n\}$, and a measurement. We assume these operations are governed by a noise mechanism such that each one of these operations corresponds to a well-known operator/superoperator;i.e., state preparation, ith-quantum gate and the measurement respectively corresponds to a state $\ket{\ket{\rho_o}}$, a quantum channel $\hat{G_i}$ and a measurement operator $\bra{\bra{M_0}}$. Therefore, as we know these operators, we can compute the expectation values of any circuit built as a combination with repetition of elements of this gate set. For instance, the measurement  applied to a circuit defined by state preparation followed by a sequence of length $l_c$ of gates  $G_{S^c_1},G_{S^c_2}...,G_{s^c_{l_c}}\  s.t \ S^c\in\{1,2...,n\}^{l_c}$  is predicted to generate the following expectation values:
\begin{equation} 
E_c=\bra{\bra{M_i}}.\hat{G}_{S^c_{l_c}}....\hat{G}_{S^c_2}.\hat{G}_{S^c_1}.\ket{\ket{\rho_i}}
 \end{equation}
Let's put ourselves in the inverse case where $\ket{\ket{\rho_o}}$, $\hat{G_i}$ and $\bra{\bra{M_0}}$ are unknown and we are given the possibility of sampling a set of N circuits $\{c_1,c_2...,c_N\}$ with corresponding sequences $\{S^{1},S^{2}...,S^{n}\}$ where $S^{1}\in \{1,2...,n\}^{l_{i}}$ . The claim of the GST protocol is that  the operators $\ket{\ket{\rho_o}}$, $\hat{G_i}$ and $\bra{\bra{M_0}}$ can be estimated based on the observed measurements  \cite{GST2q,gstintro}. In other words, if the measurement of the circuits on the target QPU generated  measurements $[\tilde{E}_{c_1},\tilde{E}_{c_2}...,\tilde{E}_{c_N}]$, the gate set $\ket{\ket{\rho_o}}$, $\hat{G_i}$ and $\bra{\bra{M_0}}$ is inferred by constructing a set of operations generating similar measurements values. This reconstruction is equivalent to the following optimization problem:

Given: a set of circuits $\{C_i\}$ and the corresponding average of observed measurements $\{E_i\}$ on target QPU:

minimize the following L1 loss function:
\begin{equation} 
fitness=\sum_i|\bra{\bra{E^{C_i}}}.\hat{G}_{S^i_{l_i}}....\hat{G}_{S^i_2}.\hat{G}_{S^i_1}.\ket{\ket{\rho^{C_i}}}-E_i|
 \end{equation}

 where $|.|$ refers  to the absolute value. This optimization is identified as a Total Variation Diminishing (TVD) problem whom the freedom degrees are the nontrivial elements of the operators describing the gate set.  The fact that the gates are quantum maps together with the Hermicity and the unit  trace of density matrices and measurement operators imply the following linear constraints on their entries):
\begin{eqnarray}
\sqrt{2}\leq \rho_{i},E_{i} &\leq&\sqrt{2}\ \ \ \ \forall\  i,j\in \{1,2,3,4\} \\\nonumber -1\leq G_{i,j} &\leq&1\ 
\end{eqnarray}

The starting point in our optimizations is the operators corresponding to the perfect target operations (no errors). As there is gauge freedom in these operators  \cite{gauge}, one should ensure that the estimations are close to the perfect operations. Therefore, we intuitively tighten these linear constraints around the perfect operation with a marge of $\pm0.1$ for gates and $\pm0.2$ for SPAM operators, and hence, given the perfect operations $\rho^p,E^p$, and $G^p$, These constraints become:

\begin{eqnarray}
max(\rho^p_i,\rho^p_i-0.2)\leq \rho_{i} &\leq& min(\rho^p_i,\rho^p_i+0.2)\ \\\nonumber max(E^p_i,E^p_i-0.2)\leq E_{i} &\leq& min(E^p_i,E^p_i+0.2)\\\nonumber max(G^p_{i,j},G^p_{i,j}-0.1)\leq G_{i,j} &\leq& min(G^p_{i,j},G^p_{i,j}+0.1)
\end{eqnarray}

\begin{figure}[h]
\centering
\includegraphics[scale=0.60]{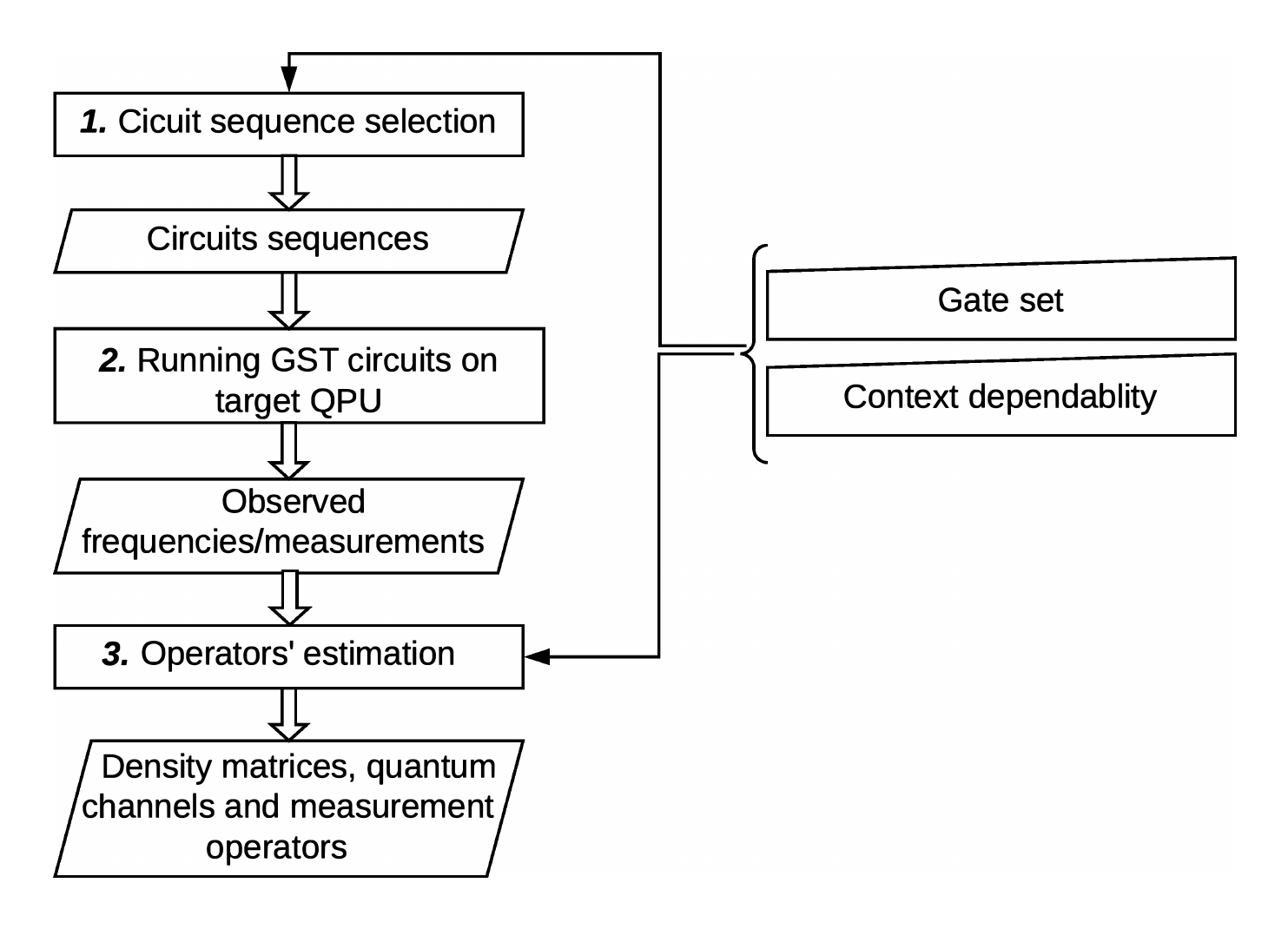}
\caption{Diagram highlighting three processing steps "rectangles" that describe our proposed version of the GST protocol: circuit sequence selection, Running the GST circuits on the target QPU and the reconstruction of the operators. The shapes of the different constituents of the diagram are defined according to flowchart convention}
\label{Norms}
\end{figure}
Furthermore, the GST relies on sampling a large number of circuit sequences to provide an accurate characterization of the operations within a target QPU. However, sampling any set of circuits sequence does not ensure an accurate characterization. In fact, gate errors can either add up or cancel out. The latter case leads to the insensitivity of the observed measurement to error parameters, and hence, alters the accuracy of the protocol. Therefore, as displayed in Figure \ref{Norms}, the first step of the GST protocol, introduced in the supplementary material, is to design a set of sequences that guarantees the full imaging of the core sequences and the amplification of the manifestations of the targeted errors as these sequences' length increases.  The second step consists of running multiple shots of the circuits designed in step 1 on the target QPU and collecting the observed measurements. Finally, the third step, as we just described in this section, is to estimate the operators describing the gate set using the observed measurements in step 2.

  In the next sections, We describe the adaptation of this protocol;i.e. the introduction of context-awareness, to infer temporal and spatial correlation in superconducting qubits and report our findings. Note that we consider it convenient to focus on errors affecting idling gates. This convenience comes from the fact that the impact of the investigated noise mechanisms is better observed/understood when affecting idling errors. Also, the accuracy of the GST protocol characterization of idling gates is very high compared to operational errors. The latter is altered by the gauge equivalence of quantum channels  \cite{gauge} and remain a challenge that we don't address in depth in this work,

\section{Crosstalk evaluation}
\begin{figure}[h]
\centering
\includegraphics[scale=0.9]{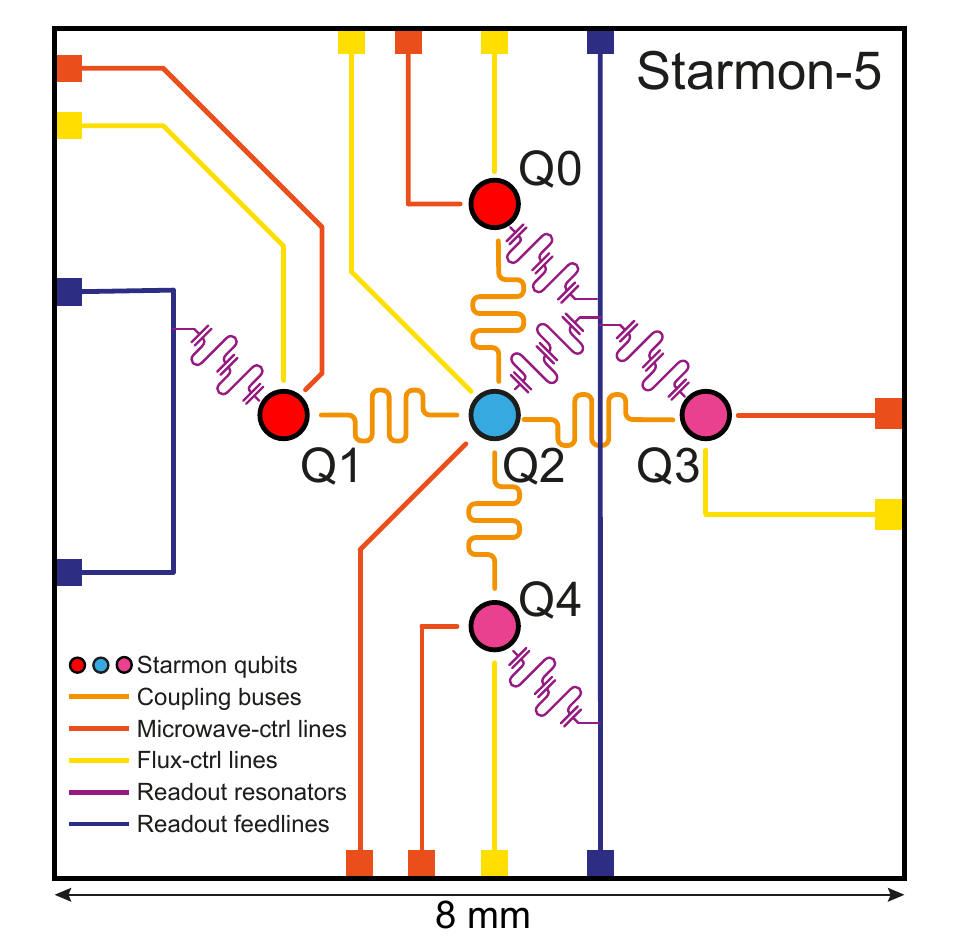}
\caption{A schematic of the Starmon-5 chip: Starmon-5 is a superconducting quantum processor based on circuit quantum electrodynamics  \cite{transmon1,transmon2}. It consists of five transmon qubits  \cite{transmon3} in an X-shaped coupling configuration  \cite{inspire}. 
Dedicated bus resonators connect nearest-neighbor pairs, enabling two-qubit gates. Each qubit is also connected to a
microwave-control line for single-qubit gating, a flux-control line for two-qubit gating, and a dispersively-coupled resonator for readout. Every readout resonator has an accompanying Purcell filter  \cite{transmon4}. The readout resonator structures for qubits Q0, Q2, Q3, and Q4 couple to one feedline, and those for qubit Q1 to another. The unique frequencies of
readout structures on a common feedline allow simultaneous, independent qubit readout by frequency multiplexing.
The high connectivity of these transmon qubits (up to 7 ports) gives them a characteristic star shape and hence the
nickname Starmon  \cite{transmon5}. All inputs and outputs to the chip are connected to a Cu printed circuit board using Al wirebonding (taken from  \cite{inspire}).}
\label{starmon5}
\end{figure}
A quantum operation on a set of qubits may induce errors to operations simultaneously applied to neighboring qubits via crosstalk effects. In current superconducting qubits, crosstalk is prominent when C-phase gates are in play  \cite{ctrb}. If we look at the geometry of a Starmon-5 chip in Figure \ref{starmon5}, we see that qubit Q1 is physically connected to Q2. Also, when a single qubit gate is applied to Q1, Q2 can undergo three different C-phases with Q0, Q3, or Q4. The goal of this subsection is to use the GST protocol to characterize the impact of these C-phases on the operations simultaneously applied to Q1. To do so, we assume that depending on the operations applied to the rest of the qubits of the chip, a quantum gate, say an idling gate $I$, can be described by multiple superoperators $G_{I|\{S\}}$. Each superoperator is specific to a context $S_i\in S$ that is defined by the operations applied to the rest of the qubits of the chips. We include contexts $S_1,\ S_2,\ S_3 $ that corresponds respectively to Q2 undergoing C-phases with Q0, Q3, and Q4 (while the two qubits left are idling) and a reference context $S_4$ where all the qubits are idling. Furthermore, to achieve informational completeness, we enable the possibility of including ancillary $R_x$ and $R_y$ rotations on Q1 in the context $S_4$ in our circuits. However, as our main goal is to infer idling gates, these ancillary operations' corresponding error generators are discarded from matrix $B$ in the germs selection algorithm. Using the genetic algorithm from the MatLab optimization toolbox, the latter provides the following set of germs: 

\begin{eqnarray}
g_{ct}&=&\{I^{1},
I^{2},
I^{3},
I^{4}R^4_x,
I^{3}I^{1},
I^{3}I^{4},
I^{1}I^{4}I^{4}I^{2},
I^{1}I^{4}I^{3}I^{4},\nonumber\\&&
I^{2}I^{3}R^4_yI^{1},
R^4_y I^{4}I^{2}I^{2}I^{1}R^4_x,
I^{2}I^{1}R^4_yI^{1}I^{3}R^4_x,\nonumber\\&&
I^{2}R^4_x R^4_yI^{3}I^{1}I^{4}I^{2}I^{3},
R^4_y I^{1}R^4_y R^4_x, I^{2}I^{2}I^{3}I^{3},\nonumber\\&&
I^{1}R^4_y  I^{1}I^{2}I^{3}I^{3} R^4_x R^4_y\}
\end{eqnarray}
Note that the upper index $i$ refers to the context   For instance, $I^{1,2,3}$ respectively describe the idling operations on Q1 that is accompanied by  Cphases on Q2 together with respectively Q0, Q3, Q4 while the rest of the qubits is idling. Therefore, each of these operations are captured by running 5-qubits gates . Since C-phases in the Starmon-5 are implemented within three times the duration of single-qubit gates, to every single-qubit gate is appended an idling time to ensure the parallelism within the elements of the gate set. Our characterization yields the following Pauli superoperators: 

\begin{equation}G(I|S_1)=\footnotesize\begin{pmatrix}
1 & 0 & 0 & 0\\
0.0000   & 0.9891  & -0.0079&    0.0007\\
   -0.0001 &   0.0084   & 0.9900 &  -0.0010\\
    0.0041  &  0.0022  & -0.0000  &  0.9959
    \end{pmatrix}
\end{equation}
\begin{equation}
G(I|S_2)=\footnotesize\begin{pmatrix}
1 & 0 & 0 & 0\\    
    0.0010  &  0.9896  & -0.0142  &  0.0003\\
   -0.0001  &  0.0152  &  0.9881  & -0.0024\\
    0.0051  &  0.0028 &   0.0015 &   0.9950\\
\end{pmatrix}
\end{equation}

\begin{equation}
G(I|S_3)=\footnotesize\begin{pmatrix}
1 & 0 & 0 & 0\\    
  0.0006   & 0.9910  & -0.0155  & -0.0010\\
   -0.0006  &  0.0164  &  0.9914 &  -0.0019\\
    0.0043  &  0.0029 &   0.0007 &   0.9962\\
\end{pmatrix}
\end{equation}
\begin{equation}
G(I|S_4)=\footnotesize\begin{pmatrix}
1 & 0 & 0 & 0\\    
   0.0005   & 0.9969 &  -0.0050  & -0.0004\\
    0.0002  & -0.0036  &  0.9970 &   0.0007\\
    0.0030  &  0.0005 &   0.0008  &  0.9964\\
\end{pmatrix}
\end{equation}

 These operators' respective diamond distances to the target perfect operators (no errors) $||G(I|S_i)-I_{perfect}||_\Diamond$ are 0.0160, 0.0218, 0.0210, and 0.0083.   The diamond distance is computed using QETLAB  \cite{qetlab} and The  CVX-package \cite{cvx,cvx2}.  Furthermore, this characterization allows us to study more features of this noise mechanism. For instance, due to the apparent anti-symmetry of the unital block with respect to the diagonal, we see a clear resemblance of these matrices to elements of the $SO(3)$ (the group of $3D$-rotations). An idling gate that displays such a feature of noise would imply that this form of noise is coherent, and hence, avoidable by inversion. To address this hypothesis, we seek for each operation $G(I|S_i)$ an unitary operator $U$ that minimises $D_i=||U_i*G(I|S_i)-I_{perfect}||_\Diamond$. Finding $U_i$ is an optimization problem in which the freedom degrees are the 3 angles that specify $U$. By solving these problems using the SQP algorithm from the Matlab optimization toolbox, we find that for scenarios 1, 2, and 3 it drops to 0.0134, 0.0148, and 0.0119. while there is no apparent improvement in context 4;i.e. $0.0083$.

 \section{Memory effects}

In this section, we address the situation where gate errors are accompanied by an error mechanism that lasts slightly longer than the implementation time of the gates themselves. In this case,  we say that the environment has a first-order memory:i.e., it alters the current gate differently depending on the previous operation. Therefore, we suppose that each gate is described by multiple superoperators. Each of these superoperators corresponds to a different context that is defined by the previous operation. Suppose we are targeting a gate set composed of an $R_x$, an $R_y$, and an idling gate, $I$. If such memory effects are displayed, each one of these gates has three corresponding superoperators. see  Table \ref{table:1}. 

\begin{table}[h!]
\centering
\begin{tabular}{|l||*{4}{c|}}\hline
\backslashbox{previous}{current}
&\makebox[3em]{$R_x$}&\makebox[3em]{$R_x$}&\makebox[3em]{$I$}
\\\hline\hline
 $R_x$   & $G_1$      & $G_4$     & $G_7$\\\hline
$R_y$  & $G_2$      & $G_5$     & $G_8$ \\\hline
$I$      & $G_3$      & $G_6$     & $G_9$ \\\hline
\end{tabular}

\caption{Superoperators representing the applied (current) gates with respect to the previously applied gates}
\label{table:1}
\end{table}

This configuration leads to constraints on the possible circuits. In fact, in order to be valid, a sequence $S=\{1,2..9\}^l$   should satisfy:

\begin{equation}
  S_i \in
    \begin{cases}
      \{1,4,7\} & S_{i-1}\in\{1,2,3\}\\
     \{2,5,8\} & S_{i-1}\in\{4,5,6\}\ \  \ \  \forall\  (1<i\leq l)\\\{3,6,9\} & S_{i-1}\in\{7,8,9\}
    \end{cases}       
\end{equation}

These constraints are injected into the germs selection algorithm and the sequences are built accordingly. As we are inferring how memory effects are affecting idling gates, the six gates $G_1$ to $G_6$ are introduced as ancillary gates, and the errors in $G_7,\ G_8$, and  $G_9$
are targeted by our germs selection algorithm. Using the genetic algorithm from the MatLab's optimization toolbox, our germs selection algorithm provides the following set:
\begin{eqnarray}g_{mem}&=&\{I^1R^f_x,
I^2R^f_y,
I^3I^f,
I^2R^3_yI^1R^f_x,
I^1R^3_xI^2R^f_y,\nonumber\\&&
R^3_yI^1R^3_xI^f,
R^2_yR^2_yR^3_yI^2R^3_yI^f,
I^2R^3_yI^1R^3_xI^2R^f_y.\nonumber\\&&
I^3I^2R^1_yR^3_xI^1R^1_xR^3_xI^f.
R^2_xR^3_yI^3I^1R^3_xI^3I^1R^f_x.\nonumber\\&&
R^3_yI^3I^2R^3_yI^3I^3I^1R^f_x\}
\end{eqnarray}
\begin{figure}[h]
\centering
\includegraphics[scale=1]{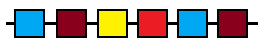}
\caption{Quantum circuit corresponding to the sequence $(3,7,6,8,3,7)$ where the blue, brown, yellow, and red gates respectively correspond to $R^3_x$, $I^1$, $R^3_y$, and $I^2$ . This sequence is an example generated by the first preparation fiducial $\varnothing$, the unrepeated (l=1) fourth germ $I^2R^3_yI^1R^f_x$ (f=3 as an ideling gate is included in every state preparation), and the second measurement fiducial $I^1R^3_x$. Note that this fiducial's first gate is in the third context as the last gate of the germ is an idling gate.   }
\label{circuitmem}
\end{figure}

where f is defined by the fiducial sequence's last gate. We also merge every state initialization with a consecutive idling gate to ensure that the context of the first gate is always 3 (preceded by an idling gate). We also include another idling gate before the final measurement so the latter is operated only in the $3^{rd}$ context (the measurement is not merged with the idling gate). Figure \ref{circuitmem} displays a circuit example. We run this designed sequence on qubit Q1 of a Starmon-5. The estimated superoperators of the three Idling gates are:

\begin{equation}G(I|S_1)=\footnotesize\begin{pmatrix}
1 & 0 & 0 & 0\\
   0.0025  &  0.9987 &  -0.0082  & -0.0105\\
    0.0047   & 0.0092  &  0.9971 & 0.0009\\
   -0.0019   & 0.0118 &    0.0062  &  0.9975
\end{pmatrix}
\end{equation}

\begin{equation}
G(I|S_2)=\footnotesize\begin{pmatrix}
1 & 0 & 0 & 0\\ 
 -0.0074 &    0.9973 &   0.0126 &  -0.0071\\
   -0.0029  & -0.0160 &    0.9994  & -0.0040\\
   -0.0066  & -0.0022  &  0.0012 &    0.9992

\end{pmatrix}
\end{equation}

\begin{equation}
G(I|S_3)=\footnotesize\begin{pmatrix}
1 & 0 & 0 & 0\\   
   -0.0001  &  0.9971  & -0.0034  &  0.0005\\
   -0.0000   & 0.0040  &  0.9966  &  0.0000\\
    0.0013  &  0.0001  & -0.0005 &  0.9987

\end{pmatrix}
\end{equation}

where $G(I|S_i)$ corresponds to the idling operator injected at the context $S_i$ that is defined by the previous gate ($S_1$, $S_2$, and $S_3$ refers to operations preceded by respectively $R_x$, $R_y$, and $I$).  These gates'  diamond distances to the ideal gates $||G_i-I||_\Diamond$  are
0.0186, 0.0265 and 0.0057. Seeking and injecting unitaries $U_i$s that minimise the distances $||U_iG_i-I||_\Diamond$ allow to respectively reduce them to 0.0156, 0.0174 and 0.0057.

\section{Discussion and outlook}

This paper introduces a context-aware version of the GST protocol;i.e., It has the ability to distinguish operations with respect to their context. This adaptation leads to a sharp characterization of errors when they incorporate such context dependencies. To do so, we described the core of our version that consists of a polynomial evaluation of the quality of the circuit sequences. As these evaluation functions admit a polynomial representation, they are computed efficiently. In fact, the fitness function of the germs selection algorithms used in this work take no more than $\sim20 sec$ in a single core of a cluster node with 2 $\times$ Xeon E5-2683 v3 CPUs (@ 2.00GHz = 28 cores / 56 threads) and 24 x 16GB DDR4 = 384GB memory. Furthermore, we evaluate the accuracy of our estimations by using simulated QPUs that fit the assumed model. This evaluation shows that our estimation is two orders of magnitudes more accurate than the gate themselves. Moreover, we adapt the GST protocol to take into account and infer spatial correlations;i.e. crosstalk errors. The implementation of the designed protocol on the Starmon-5 chip shows that when a C-phase is applied to neighboring qubits, errors are induced to the studied qubits via crosstalk. Fortunately, a large fraction of these errors is reversible. For instance, $1-0.0119/0.0210 \sim 43\%$ of the errors in Q1's idling when a C-phase is applied to qubits Q2 and Q4 are coherent. This offers the opportunity of drastically mitigating the errors that are purely due to crosstalk;i.e. in this example a reduction reaching a factor of $(0.0210-0.0087)/(0.0119-0.0087)\sim 4 $  is predicted. Next, we addressed the problem of temporal correlation, namely memory effects. This study explicitly shows that when an idling gate is applied after a rotation gate, errors due to memory effects are induced. However, up to $32\%$ of these errors is reversible which constitutes a reduction of the errors purely due to memory effects by a factor of  $\sim 2$. Note that measuring the sequences for these characterizations (designed for a maximal repetition index 6) took respectively $30$ and $24$ hours for the crosstalk and memory effects experiments.

To conclude, this work takes a step further the agreement between the assumed noise model by GST protocol and real-life error mechanisms. However, the efficiency of the hereby proposed Context-Aware GST (CA-GST) protocol is altered by sampling errors and the presence of errors mechanisms that reside beyond the assumed models. Nevertheless, due to their hereby demonstrated coherent aspect, this work promises a large reduction in some of the most prominent error mechanisms in this era.

\begin{acknowledgments}
The authors would like to thank Diogo Valada, and Hans van Someren for their technical support and Miguel Moreira for describing to us the operations on Quantum-Inspires's Starmon-5 chip. The authors would like to acknowledge funding from Intel Corporation.
\end{acknowledgments}

\bibliography{refs.bib}
\maketitle

\begin{center}
\textbf{\large Supplemental Materials: A context-aware gate set tomography characterization\\ of superconducting qubits}
\end{center}
\setcounter{equation}{0}
\setcounter{figure}{0}
\setcounter{table}{0}
\setcounter{page}{1}

\makeatletter
\renewcommand{\theequation}{S\arabic{equation}}
\renewcommand{\thefigure}{S\arabic{figure}}
\renewcommand{\bibnumfmt}[1]{[S#1]}
\setcounter{section}{0}
\renewcommand{\thesection}{S-\Roman{section}}
\section{Circuits layout}

\begin{figure}[h]
\includegraphics[scale=0.5]{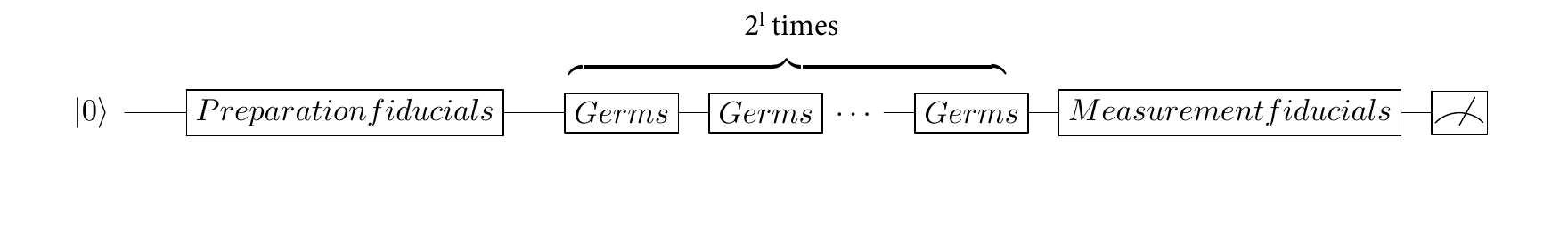}
\caption{GST sequences composition. Each sequence is composed of a state preparation in the ground state followed by germs "sandwiched" by a set of preparation and measurement fiducials followed a measurement in the computational basis at the end. }
\label{layout}
\end{figure}

The GST protocol is based on sampling a predetermined set of circuits. These circuits are tailored with the aspiration of triggering and observing errors within a gate set. A gate set is composed of a set of state preparations, a set of gates, and a set of measurements.  To enable the reconstruction of the underlying operations, the sets of state preparations and measurements are required to satisfy  Informationally Completeness (IC);i.e., a set of $2^{2n_q}$ ($n_q$ being the number of qubits)  operators spanning the space of linear operators that act on the $2^{n_q}$ dimensional Hilbert space  \cite{ICpovm,pygstynew}. However, most of the qubit technologies, including the one that we are targeting in this work, inherently allow the preparation of the ground state and the measurement on the computational basis (whether or not in the excited state). Therefore, as illustrated in Figure \ref{layout}, in lieu of applying gate sequences (germs) to an IC set of prepared states and measuring in IC  set of measurement bases, each circuit is commenced by a state preparation in the ground state followed by a preparation fiducial (a gate sequence with length $\leq 3$) and terminated by a measurement fiducial followed by a measurement in the computational basis.  Injecting the right set of preparation and measurement fiducials is equivalent to the preparation and the measurement in an IC set of states and measurements.   Furthermore, the germs should guarantee an all-inclusive manifestation of the target error mechanisms within the gate set elements and this manifestation is expected to come with a higher amplitude as the length of germs grows. In the following, we introduce our selection algorithms allowing us to design efficient fiducial and germ sequences.  

\section{Fiducials selection}

In the density matrix and quantum channels formalism, the expectation values of the measurement in a  circuit having the form displayed in Figure \ref{layout} can be expressed as:
\begin{equation} 
E = \bra{\bra{M}}|f_m.G.f_p\ket{\ket{\rho_0}}
 \end{equation}
 
Where $\ket{\ket{\rho_0}}$  corresponds to the density matrix of the prepared state, $f_p$ is the preparation fiducial's superoperator, $G$ is the superoperator representing the sequence of $2^l$ of germs, $f_m$ is the measurement fiducial's superoperator and $\bra{\bra{M}}$ is the measurement operator. The intuition behind the introduction of the preparation and the measurement fiducials is to maximize the sensitivity of the observed measurements to every nontrivial entry of the superoperator $G$. By looking at  Equation 1, we see that $E$  is a first degree polynomial of these entries and therefore:

\begin{equation} 
E=\sum_{i=2,j=1}^{i,j=4}a_{i,j}G_{i,j}+a_0 
 \end{equation}
Therefore, the ability to sense the elements of the superoperator G in the measurement observed using a set of k preparation and k measurement fiducials are represented in the following vector:
\begin{equation} 
T=vect\big{(}\sum_{p,m=1}^{k}|a_{i,j}^{p,m}|\big{)}
 \end{equation}
where the upper index refers to the circuit defined by a preparation fiducial $p$ and a measurement fiducial $m$. Our fiducial selection algorithm is the following optimization problem:

given a set of all possible fiducials, maximize:
\begin{equation} 
 fitness=sum(T)/var(T)
 \end{equation}

The freedom degrees of this optimization problem are the combinations of at most three elements of the gate defining each element of the set of fiducials. To ensure designing informationally over-complete sets of fiducial for single-qubit cases, 6 different fiducials are injected in both sides (preparation and measurement).  Note that, especially when considering a smaller set of fiducials, one should make sure that all the elements of $T$ are nonvanishing.

\section{Germs selection}

 Tailoring a set of circuits that ensures the manifestation of the errors within the gate set is key in the GST protocol. To address this problem,  we assume that each gate $G_i$ in the gate set is expressed as its corresponding perfect gate $G^{p}_i$ preceded by an erroneous part expressed in the exponential form. This separation is expressed in the matrix multiplication form as:

\begin{equation} 
G_i=G^{p}_i e^{\mathcal{L}_i}
 \end{equation}
These error generators $\mathcal{L}_i$ vanish for perfect gates and get larger absolute values when the gate is noisier. Therefore, for gates within the moderate accuracy range, the previous equation admits the following Taylor expansion:
\begin{equation} 
G_i\simeq G^p_i(\mathcal{I}+\mathcal{L}_i+\Theta(\mathcal{ L}_i))
 \end{equation}
$\mathcal{I}$ being the identity operator. The manifestation of  errors in the observed measurements is related to the dependency of the measurements on the components of these error generators. In fact, since germ $G$ and fiducials $f_m$ and $f_p$ in Equation 1 are built as combinations of gates within gate set $\{\ket{\ket{\rho_o}},G_1,...,G_n,\bra{\bra{M}}\}$, the expectation value $E_c$ is a polynomial with respect to the components of the error generators $\mathcal{L}_i$  that characterise each gate $G_i$ in the gate set. Storing and processing such polynomials is expensive especially for long depth circuits. Therefore, we track the manifestation of the entry of indices $(j,k)$ of the error generator $\mathcal{L}_i$ (;i.e.,  ($\mathcal{L}_{i\ (j,K))}$) in measurement $E^c$ in a circuit $c$ using the simplified quantity
\begin{eqnarray}
 E^c\bigg{|}^{\mathcal{L}_{a(b,c)}=0\ (a,b,c)\neq(i,j,k)}_{\mathcal{L}_{a(b,c)}=x\ (a,b,c)=(i,j,k)} =
a^c_{0,i,j,k}+a^c_{1,i,j,k}x+\footnotesize{\Theta }
\end{eqnarray}

We see that variable $a^c_{1,i,j,k}$ represents the magnitude of the impact  on measurement $E^c$ by entry $(j,k)$ of the error generator $\mathcal{L}_{i}$ characterizing the gate element $G_i$. Therefore, for a set of $N$ circuit, we track the impact of the ensemble of the error generators' entries on the outcome of the measurement using the following vector
\begin{equation} 
 A=vec\big{(}\sum_{c}|a_{1,i,j,k}^c|\big{)}
 \end{equation}

In other words, each entry in $A$ represents a single entry of the error generator of one of the gates. For each $n$-qubit gate in the gate set, the size of $A$ is increased by $4^n(4^n-1)$ (which is the number of nontrivial elements in the errors generator).  Note that each circuit's index  $c$ is defined by the indices of the SPAM fiducials $p$ and $m$, the germ index $g$, and the repetition index $l$. The repetition of the germs is introduced as the GST protocol is based on sampling a large set of circuits on which an exhaustive search is impractical. However, this simplification reduces the search space of all possible sequences for such a large set of circuits. Fortunately,  in the GST protocol is considered a small subset of circuits with a short depth called germs, and, as illustrated by Figure \ref{layout},   large circuits are built by repeating the germs such that $G^l=G^{l-1}G^{l-1}$. Furthermore, to ensure that the manifestation of the errors is amplified as the $l$ increases, we store each length $l$ in a distinct column in a matrix $B$.

\begin{equation} 
 B^l=vec\big{(}\sum_{p,m,g}|a_{1,i,j,k}^{p,m,g,l}|\big{)}
 \end{equation}
The matrix B is a juxtaposition of vectors $B^l$ ($1\leq l \leq L$). In each entry of these vectors is accumulated the amplitude of the polynomial factors in the ensemble of circuits with repetition index $l$ of the entry$(i.j)$ of a gate $k$.  Our germs selection algorithm is the following:

given the set of all possible germs, maximize:
\begin{equation} 
fitness=Min(B^{L}).
 \end{equation}
Where $Min(B^{L})$ refers to the smallest value of the vector accumulating the entries of the error generators within circuits with maximal repetition index $L$. The freedom degrees are the combination that defines the set of germs. The fact that the errors should be amplified as $l$ increases is introduced as the following constraint:

\begin{eqnarray}
B_i^l&<&B_i^{l+1} \\  \forall\  \  1\leq l\leq L-1\ , \ &1&\leq i\leq \sum_{k=1}^{||G_k||}4^{n_k}(4^{n_k}-1) \nonumber
\end{eqnarray} 
  with $n_k$ being the number of qubits on which the gate $G_k$ acts and $||G_k||$  the number of gates in the gate set. Therefore, maximizing the fitness function enables obtaining circuits with a large accumulation of errors. Furthermore, the constraints ensure the growth of the this accumulation with the repetition index $l$.
\section{Accuracy evaluation}
\begin{figure}[h]
\centering
\includegraphics[scale=0.67]{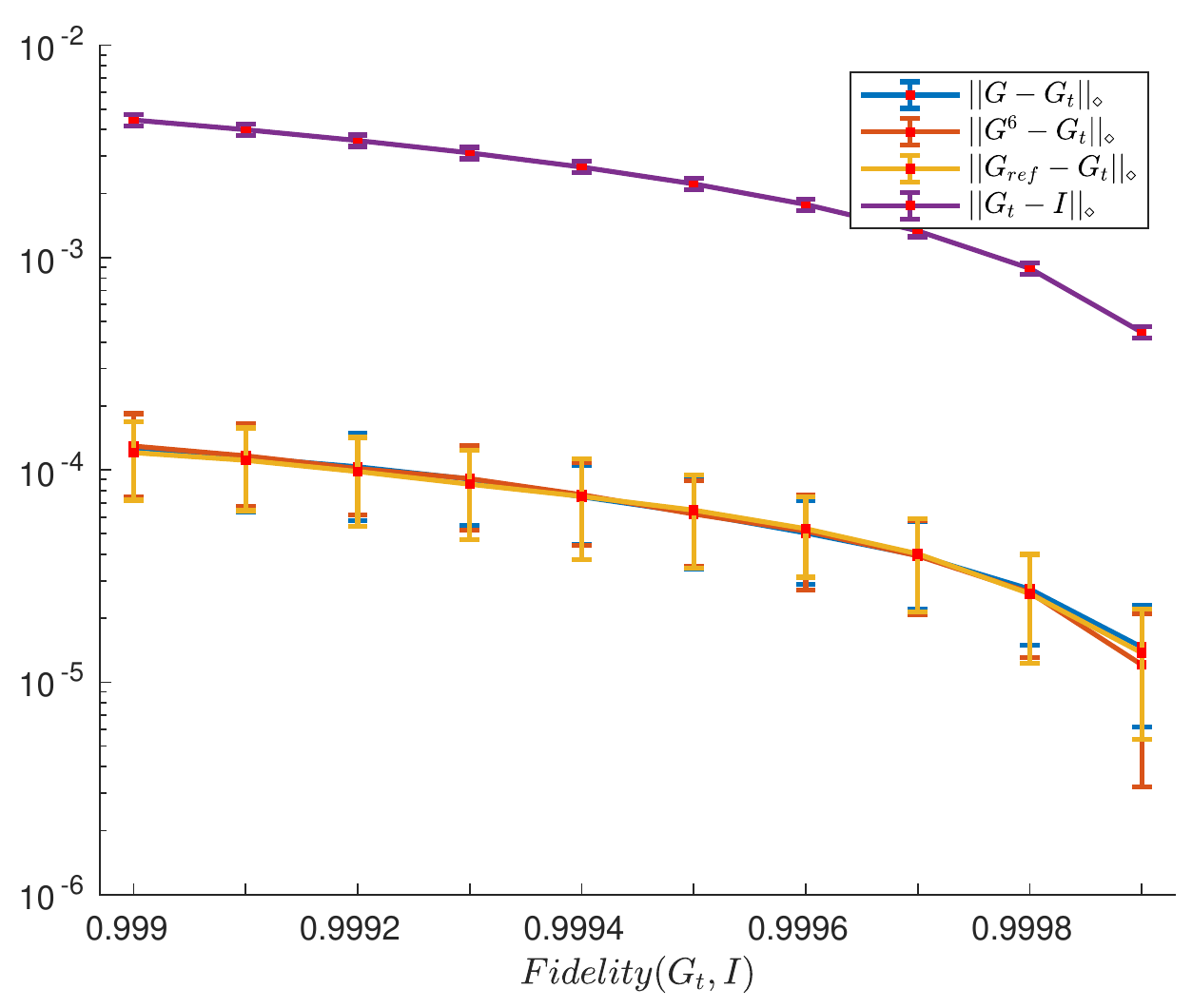}
\caption{Achieved accuracy in characterizing idling errors using our protocol and the reference protocol. The blue, red, and yellow lines represent the averaged discrepancy between and 100 randomly generated target gate sets $G_t$ and their GST estimations using the gate sets $g$, $g^6$ and $g_ref$.  The violet line represents the averaged distance between the $g_t$ and the perfect idling gate. The distances are measured with respect to the diamond norm. Note that in our simulations, a QPU is defined by a gate set. Within this gate set, SPAM operators are defined by a density matrix and a measurement operator having fidelities within a reasonable range. Furthermore, each gate $G$ is generated as a convex sum of $99.9\%$ portion of its corresponding perfect gate ( no errors) $G_p$ and another noisy $0.1\%$ portion generated as a random channel. As proposed in  \cite{UMC}, the fidelity of these gates is controlled by linearly tuning their error generator $\mathcal{L}$ defined by $G=G_pe^\mathcal{L}$. In this figure, each data point corresponds to an averaging over a $100$ distinct gate set.}
\label{compare}
\end{figure}

To evaluate the accuracy of our circuit sequence selection algorithms, we compare our protocol's accuracy with its commonly known counterpart \cite{pygstynew}. As the latter was designed to perform the standard GST (agnostic to context-dependent errors), our benchmarks will be restricted on comparing the ability to infer static errors. The target devices of our proposed experiments are a set of simulated QPUs with various known error generators. Due to our knowledge of the operators describing the simulated operations, it is possible to compare the output of both protocols with respect to these operators. In these experiments, we consider gate sets composed of a state preparation in the ground state, two rotation gates $R_x$ and $R_y$ of an angle $\pi/2$ around the axes X and Y, an idling gate $I$, and measurements in the computational basis. This particular gate set was targeted using the reference protocol where the selected circuits were described by the following sequence  \cite{pygstynew}:
\begin{eqnarray}
f_{ref}&=&\{ \varnothing,R_x,R_y,R_x R_x,R_x R_x R_x,R_y R_y R_y \}\\
g_{ref}&=&\{R_x,R_y,I,R_x R_y,R_x R_y I,R_x I R_y, R_x I I,R_y I I,\nonumber\\&& R_x R_x I R_y, R_x R_y R_y I, R_x,R_x R_x R_y R_x R_y R_y\} 
\end{eqnarray}

where $f_{ref}$ and $g_{ref}$ respectively correspond to germs and fiducial sequences, $\varnothing$ the empty element (no gate). The repetition(duplicating the germ as in Fig \ref{layout}) in the reference protocol is done until the germ sequences reach 256 gates. We introduce the same gate set in our circuit sequence algorithms without targeting any context dependency and the germs selection algorithm yield the following set of germs.

\begin{eqnarray}
g&=&\{R_x,R_y,I,R_yR_x,R_yR_y,II,R_yR_xI,IR_yR_x,\nonumber\\ &&IR_yR_xR_y,IR_yIR_x,R_yR_xIR_x \} 
\end{eqnarray}
To ensure having approximately the same number of sequences with the reference protocol, this set of germs is designed for a maximal repetition index of 7. We also design a set of germs $g^6$ that is designed for a maximal repetition index of 6 with longer initial sizes of germs which is closer to the cases we address in the main text. The obtained $G^6$ is the following:
\begin{eqnarray}
g^6&=&\{II,
IR_x
,IR_y
,R_yIR_yR_x
,IR_xR_yR_y,
IR_xR_yR_x,\nonumber\\ &&
R_yR_yR_xIIR_y,
IR_xR_yIR_yI,
IR_xIR_xIIIR_y,\nonumber\\ &&
IR_xR_yIR_xIII,
R_yIR_xIR_yIIR_x\}
\end{eqnarray}
These optimization problems were solved using the genetic algorithm from MatLab's optimization toolbox with 10 stall generations.  We use set $f_{ref}$ as it maximizes our fiducial sequences selection algorithm's fitness function. In our comparison, we use the three sets of sequences to characterize sets of simulated QPUs with gates $G_I, G_{X90}$ and $ G_{Y90}$ incorporating random errors yet with fixed fidelities. Assuming that our protocol produces $\tilde{G}_I, \tilde{G}_{X90}$ and $ \tilde{G}_{Y90}$, and the estimation using the reference circuits produces $\tilde{G}^{ref}_I, \tilde{G}^{ref}_{X90}$ and $ \tilde{G}^{ref}_{Y90}$. We compare the quality of the sequence selection algorithms with the reference by comparing the distinguishability between $\tilde{G}_I$ and $G_I$  and between $\tilde{G}^{ref}_I$ and $G_I$. The same approach is applied to evaluate $\tilde{G}^6_i$ (the superoperator obtained using $g^6$). Figure \ref{compare} displays these measures with respect to the diamond norm.

In Figure \ref{compare}, we see that the use of our sequence selection derived circuits leads to a similar accuracy (blue and orange lines) when using the standard circuits (yellow lines). For idling errors, this characterization is highly reliable as its inaccuracy represents $3\%$ of the inaccuracy of the inferred operations themselves (purple lines).
\end{document}